\documentclass[twocolumn]{aastex631}
\usepackage{amsmath}

\usepackage[normalem]{ulem}

\newcommand{\simproj}[1]{\textsc{#1}}
\newcommand{\simcode}[1]{\textsc{#1}}
\newcommand{\apo}{\simproj{Apostle}}
\newcommand{\aur}{\simproj{Auriga}}
\newcommand{\gadget}{\simcode{Gadget-3}}
\newcommand{\arepo}{\simcode{Arepo}}
\newcommand{\sersic}{S\'ersic}
\newcommand{\sech}{$\text{sech}^2$}

\newcommand{\figref}[1]{Fig. \ref{#1}}
\renewcommand{\eqref}[1]{Eq. \ref{#1}}
\newcommand{\tabref}[1]{Table \ref{#1}}

\shortauthors{Hu et al.}

\begin{document}


\title{APOSTLE vs. AURIGA Simulations: How Subgrid Models Shape Milky Way Analogs}

\author[0009-0008-9232-6337]{Jianhong Hu}
\affiliation{School of Physics and Laboratory of Zhongyuan Light, Zhengzhou University, Zhengzhou 450001, China}
\email{jhu24@zzu.edu.cn, hyang@nao.cas.cn, lgao@bnu.edu.cn}

\author[0000-0003-3279-0134]{Hang Yang}\thanks{ Corresponding author}
\affiliation{School of Physics and Laboratory of Zhongyuan Light, Zhengzhou University, Zhengzhou 450001, China}
\affiliation{Institute for Frontiers in Astronomy and Astrophysics, Beijing Normal University, Beijing 102206, China}
\affiliation{School of Physics and Astronomy, Beijing Normal University, Beijing 100875, China}

\author[0009-0006-3885-9728]{Liang Gao}
\affiliation{School of Physics and Laboratory of Zhongyuan Light, Zhengzhou University, Zhengzhou 450001, China}
\affiliation{Institute for Frontiers in Astronomy and Astrophysics, Beijing Normal University, Beijing 102206, China}
\affiliation{School of Physics and Astronomy, Beijing Normal University, Beijing 100875, China}

\begin{abstract}
 
Despite significant progress in cosmological simulations of galaxy formation, the role of subgrid physics in shaping the detailed properties of galaxies remains incompletely understood. In this work, we analyze two sets of zoom-in simulations that share identical initial conditions but adopt distinct implementations of baryonic physics, enabling a controlled comparison of their predictions. We examine the stellar properties, morphological structures, and satellite populations of the simulated galaxies at $z=0$.
We find that {\aur} galaxies systematically exhibit higher stellar masses and surface densities than their {\apo} counterparts. These differences are primarily driven by variations in the efficiency of gas cooling from the circumgalactic medium (CGM) into the star-forming gas. Both simulations form well-defined disk galaxies; however, {\aur} systems generally display higher disk-to-total mass ratios, earlier disk formation, and more prominent dynamical structures such as bars and spiral arms. Nevertheless, strongly disk-dominated systems are present in both simulations, although they do not arise in the same host haloes.
The vertical disk structure in both simulations is well described by a {\sech} density profile, with scale heights below $\sim1$ kpc in the inner regions. The satellite populations also differ, with {\aur} producing systematically more massive satellites, including a $\sim0.3$ dex increase in the most massive system, while the number of satellites above $10^6 M_\odot$ remains comparable in most halo pairs. Both simulations reproduce similar satellite stellar mass–metallicity relations, albeit $\sim0.25$ dex higher than observation.
This comparative study therefore provides useful benchmarks for future efforts to better constrain galaxy formation models.

\end{abstract}



\section{Introduction} \label{sec:intro}

The standard $\Lambda$ cold dark matter ($\Lambda$CDM) paradigm has become the cornerstone of modern cosmology, providing a robust framework for understanding cosmic structure formation \citep[e.g.][]{peebles1982, blumenthal1984, davis1985}. This model has demonstrated remarkable success in explaining diverse observations, particularly the statistical properties of large-scale structure \citep[e.g.][]{mo2002, kravtsov2004, springel2005nat}. Within this framework, galaxies form through the cooling and condensation of gas within hierarchically assembled dark matter haloes, while their luminous components are shaped by complex baryonic processes operating across a wide range of scales \citep[e.g.][]{white1978, white1991}. The nonlinear coupling of processes like gas cooling, star formation, and feedback renders galaxy formation one of the most challenging problems in modern astrophysics, motivating the use of numerical simulations to study this complex system \citep[e.g.][]{2015ARA&A..53...51S, 2017ARA&A..55...59N, 2023ARA&A..61..473C}.

Although cosmological simulations are now essential tools for studying galaxy formation, the accurate modeling of baryonic physics remains one of the most formidable challenges, primarily due to its inherently multiscale nature spanning from galactic to interstellar scales. The importance of feedback processes has been established over decades of numerical work. Early simulations showed that, without valid feedback, gas undergoes catastrophic cooling, producing galaxies that are overly compact, with excessive central star formation and significant angular momentum loss \citep[e.g.][]{navarro1991, navarro1995, navarro1997}.
Later studies demonstrated that suitably calibrated feedback—such as supernova-driven outflows can suppress runaway cooling, regulate star formation, and preserve extended gas reservoirs required for disk formation \citep[e.g.][]{weil1998, sommerlarsen1999, thacker2000, maller2002, 2014MNRAS.437.1750M}. Comparative projects using identical initial conditions across different codes, such as Aquila \citep{scannapieco2012} and AGORA \citep{kim2014, rocafabrega2021, 2025ApJ...994..245J}, have further shown that differences in subgrid physics—especially feedback prescriptions—are the dominant source of variation in hydrodynamical simulation results.

Building upon these foundations, we conduct a systematic comparison of galaxy formation outcomes between the {\aur} \citep{grand2017} and {\apo} \citep{sawala2016, fattahi2016} simulation suites to establish how variations in baryonic physics models translate into systematic differences in galaxy evolution and observable properties. The original versions of these simulations have been widely used to study galaxy formation across a range of environments, including detailed investigations of Milky Way-like systems \citep[e.g.][]{2019MNRAS.490.5182L, 2020MNRAS.497.4459F,  2025ApJ...994..271Y, 2025MNRAS.542.2443R, 2026MNRAS.545f1551O}. However, a systematic comparison of these simulation suites under matched initial conditions has not yet been performed, limiting our ability to isolate the impact of different baryonic physics implementations. In this work, we analyze a set of new simulations that adopt identical initial conditions for Local Group-like volumes, while otherwise retaining the full subgrid physics implementations of the original {\aur} and {\apo} models, particularly for stellar feedback. Specifically, {\apo} adopts a thermal energy injection scheme that heats gas particles to a fixed temperature \citep{dvecchia2012}, whereas {\aur} employs a kinetic wind model in which particles are temporarily decoupled and launched isotropically, depositing both thermal and kinetic energy into the interstellar medium upon recoupling \citep{vogelsberger2013}.

Previous studies utilizing these controlled simulation datasets have already revealed significant differences in baryonic properties arising from these distinct implementations. \citet{kelly2022} showed that the efficient stellar feedback in {\apo} leads to substantial baryon deficiency due to strong gas ejection and suppressed accretion at high redshift. In addition, \citet{yang2024} examined the evolution of angular momentum, finding that the  feedback in {\apo} inhibits the recycled galactic scale fountain process, thereby limiting angular momentum acquisition from the circumgalactic medium. Building on these results, we present a systematic and unified analysis of galaxy properties across both simulations. In particular, we examine stellar population distributions, galaxy morphology, and satellite system properties to establish a coherent physical picture of how baryonic physics implementations shape the Milky Way analogs.

This paper is organized as follows. Section~\ref{sec:samps} describes the simulation suites and the galaxy sample selection, following the methodology of \citet{yang2024}. Section~\ref{sec:res} presents our comparative analysis, including stellar population properties (Section~\ref{subsec:props_star}), morphological characteristics (Section~\ref{subsec:morph}), and satellite system properties (Section~\ref{subsec:satellites}). We summarize our findings and discuss their implications in Section~\ref{sec:discus}.



\section{Simulations and Sample Selection} \label{sec:samps}

Our samples are selected from two suites of `zoom-in' simulations with identical initial conditions but distinct hydrodynamical schemes. The first one is {\apo} project \citep{sawala2016, fattahi2016}, which was run using a modified version of {\gadget} code \citep{springel2005mn}, with fluid properties calculated by the pressure–entropy formulation of smoothed particle hydrodynamics (SPH) technique \citep{hopkins2013}. The second one is {\aur}-like simulations (with the same galaxy formation model as {\aur} project \citep{grand2017}), using moving-mesh magneto-hydrodynamics code {\arepo} \citep{springel2010}. All simulations are performed with the WMAP-7 cosmological parameters \citep{komatsu2011}: $\Omega_{\Lambda}=0.728$, $\Omega_m=0.272$, $\Omega_b=0.0455$, $h=0.704$, $\sigma_8=0.81$, and $n_s=0.967$. The simulations share identical mass resolutions, with initial gas (dark matter) particle/cell masses of about $1.2~(5.9) \times 10^5 M_\odot$ and a common maximum gravitational softening length of 307 pc.

Critically, two simulations incorporate markedly different subgrid models for key baryonic processes. Most notably, {\apo} implements a thermal feedback scheme \citep{dvecchia2012} where supernova energy is stochastically distributed among neighboring gas particles as pure thermal energy, heating them by a prescribed temperature increase. In contrast, {\aur} utilizes a kinetic feedback model \citep{vogelsberger2013} that launches isotropic `wind particles' carrying mass, metals, and both thermal and kinetic energy; these particles deposit their contents upon recoupling with the interstellar medium after reaching a low-density region or traveling for a specified time. While other factors like AGN feedback and hydrodynamical solvers may contribute to differences, previous studies suggest that stellar feedback implementations play a dominant role in shaping galaxy evolution outcomes, particularly in regulating baryon fractions and angular momentum distributions \citep{kelly2022, yang2024}. Further details on the subgrid physics are provided in \citet{schaye2015} (for {\apo}) and \citet{grand2017} (for {\aur}).

Both simulations have identical initial conditions, `zoom-in' of two Local Group-like volumes, namely S5 and V1. Our sample set comprises the first and second most massive disk galaxies of each volume in both simulations, tagged as XX-1, XX-2 respectively, where XX denotes the volume (either S5 or V1). Here the disk galaxies are identified using the kinematic rotation criterion $\kappa_{\rm rot} > 0.5$ \citep{yang2023}, where $\kappa_{\rm rot} \equiv K_{\rm rot}/K$ quantifies the fraction of kinetic energy in ordered rotation \citep{sales2012}, calculated as:
\begin{equation} \label{equ:kappa}
    \kappa_{\rm rot} = \frac{K_{\rm rot}}{K}
           = \frac{\sum_i{(1/2)}m_i[{\boldsymbol{v}_i} \cdot \hat{\boldsymbol{r}}^{\phi}_i]^2}
                {\sum_i{(1/2)}m_i{v_i^2}},
\end{equation}
where $\boldsymbol{v}_i$ and $m_i$ represent the velocity vector and mass of stellar particle $i$, $\hat{\boldsymbol{r}}^{\phi}_i = \widehat{\boldsymbol{L} \times \boldsymbol{r}_i}$ is the azimuthal unit vector in cylindrical coordinates aligned with the total stellar angular momentum $\boldsymbol{L}$. These galaxies are then matched across simulations by comparing their virial radii and centers, with full details of the sample construction available in \citet{yang2024}. The systematic pairing ensures a controlled comparison of galaxies evolving under different subgrid physics implementations but in cosmologically analogous environments.

\tabref{tab:props} lists properties of galaxies at $z=0$. While the sample galaxies show consistent virial masses ($M_{200} \sim 10^{12} M_\odot$) with minimal variation ($< 0.1$ dex) between {\apo} and {\aur} counterparts, they exhibit striking disparities in stellar content. Most notably, {\aur} galaxies exhibit systematically higher stellar masses (by $\sim$ 0.4 dex) and enhanced specific angular momentum compared to their {\apo} analogs, highlighting the substantial impact of differing subgrid physics implementations on baryonic evolution within similar dark matter halos, as discussed in e.g. \citet{kelly2022, yang2024}.

\figref{fig:projmap} demonstrates face-on and edge-on projections of our galaxies. The $z$-axis is determined as the shortest principal axis by calculating eigenvectors of the moment of inertia tensor of star particles within $3 R_{\rm hsm}$ ($R_{\rm hsm}$: half stellar mass radius). Upon visual inspection, the {\aur} galaxies exhibit more distinct substructures (particularly spirals and bars), compared to their counterparts in {\apo} simulations.

\begin{table*}
    \caption{The properties of the matched galaxies at $z=0$. From left to right, the columns are: (1) Galaxy name, with prefix Ap-/Au- for {\apo}/{\aur} simulations; (2)-(3) Galaxy virial mass and virial radius; (4) Galaxy stellar mass within projected radius 20 kpc and vertical height 5 kpc; (5) Half stellar mass radius; (6) Kinematic rotation parameter within $3 R_{\rm hsm}$; (7) Stellar specific angular momentum within $3 R_{\rm hsm}$; (8)-(10) Parameters from single {\sersic} fit.}

    \begin{tabular}{l | cc | cccc | ccc}
        \hline
            & \multicolumn{2}{c|}{Dark matter} & \multicolumn{4}{c|}{Stellar} & \multicolumn{3}{c}{Sersic fit} \\
        Name & $\log_{10}M_{200}$ & $R_{200}$ & $\log_{10}M_*$ & $R_{\rm hsm}$ & $\kappa_{\rm rot}$ & $j_*$ & $\log_{10}M_*$ & $R_e$ & $n$ \\
             & [$M_\odot$] & [kpc] & [$M_\odot$] & [kpc] & & [kpc km/s] & [$M_\odot$] & [kpc] & \\
        \hline
        Ap-S5-1 & 11.96 & 199 &  9.78 & 4.22 & 0.59 &  121 & 10.27 & 3.99 & 1.82  \\
        Au-S5-1 & 12.02 & 208 & 10.23 & 3.15 & 0.58 &  483 & 10.71 & 3.08 & 2.99  \\
        \hline
        Ap-S5-2 & 11.89 & 189 &  9.66 & 5.51 & 0.57 &  443 & 10.18 & 4.53 & 2.12  \\
        Au-S5-2 & 11.91 & 192 & 10.07 & 4.69 & 0.67 &  613 & 10.55 & 3.77 & 1.36  \\
        \hline
        Ap-V1-1 & 12.20 & 239 & 10.13 & 7.44 & 0.68 & 1076 & 10.64 & 5.72 & 1.14  \\
        Au-V1-1 & 12.20 & 239 & 10.48 & 6.14 & 0.59 & 1132 & 11.01 & 5.21 & 2.23  \\
        \hline
        Ap-V1-2 & 11.89 & 189 &  9.74 & 4.12 & 0.57 &   51 & 10.24 & 3.44 & 1.50  \\
        Au-V1-2 & 11.92 & 193 & 10.03 & 7.92 & 0.70 & 1039 & 10.60 & 7.73 & 2.24  \\
        \hline
    \end{tabular}
    \label{tab:props}
\end{table*}

\section{Results} \label{sec:res}

\subsection{Mass Assembly Histories} \label{subsec:mah}

\begin{figure}
    \includegraphics[width=0.5\textwidth]{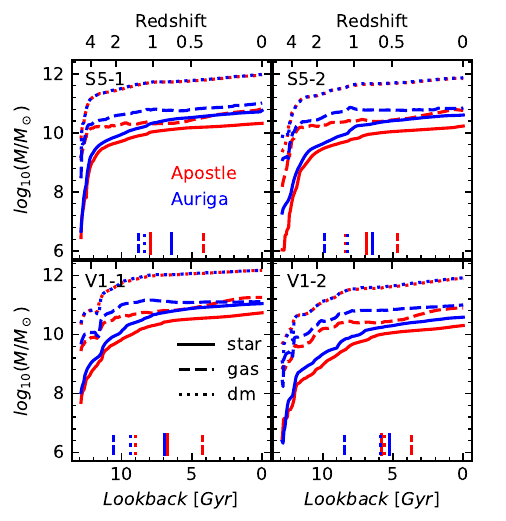}
    \centering
    \caption{Mass evolution histories of galaxies. Four representative pairs are shown (as labeled in the upper-left corners of panels), with {\apo} and {\aur} simulations indicated by red and blue curves respectively. Stellar (solid), gas (dashed), and dark matter (dotted) components are displayed separately.}
    \label{fig:evol_mass}
\end{figure}

The mass assembly histories offer the key insight into how galaxies form and evolve over cosmic time. \figref{fig:evol_mass} compares the mass assembly histories of stellar (solid), gaseous (dashed), and dark matter (dotted) components for four representative galaxy pairs from the {\apo} (red) and {\aur} (blue) simulations. Each panel is labeled with the corresponding simulation name in the upper-left corner. We also mark the half mass formation times $t_{50}$ of each component with vertical lines matching their respective line styles.

As shown in \figref{fig:evol_mass}, the dark matter components exhibit remarkably consistent assembly histories between the two simulations, with the difference in $t_{50,\rm dm}$ being less than 0.01 dex. This agreement is expected, given that both simulations adopt identical initial conditions and the same gravitational solver. However, the assembly histories of stellar components reveal systematic differences: {\aur} galaxies maintain higher stellar masses than their {\apo} counterparts from very early epochs, with the difference gradually increasing over time to reach $\sim 0.4$ dex by $z=0$. In the {\aur} simulations, enhanced pristine gas inflows and galactic scale recycling flows \citep[e.g.][]{kelly2022,yang2024} maintain a more continuous cold gas supply, which in turn leads to more extended star formation and a higher final stellar mass. Consequently, galaxies in the {\aur} have a later mean star formation time, the difference in $t_{50,*}~\sim$ 0.04 dex. The total gaseous components exhibit a self-regulated behavior: {\aur} galaxies contain significantly more total gas at high redshifts (before $\sim 10$ Gyr, with $z \gtrsim 2$), but this difference gradually decreases toward z=0.

To quantify the origin of the differences in the total stellar mass of corresponding galaxies in the two simulations, we decompose the stellar mass assembly into a sequence of mass conversion processes by tracing all gas elements that have been accreted into the halo. The stellar mass at $z=0$ can then be expressed as
\begin{equation}
M_{*} = f_{\rm ISM}\, f_{\rm re}\, f_{*}\, M_{\rm acc},
\end{equation}
where $M_{\rm acc}$ denotes the total gas mass accreted into the halo over its history. 
The quantity $f_{\rm ISM} = M_{\rm cool,all}/M_{\rm acc}$ is the fraction of accreted gas that eventually cools and enters the interstellar medium (ISM). 
A gas element is considered to have entered the ISM if it satisfies the star formation criterion (star formation rate, SFR $>0$) at least once. 
In our calculation, each gas element is counted only once, regardless of any subsequent recycling events. 
The factor $f_{\rm re} = (M_* + M_{\rm cool})/M_{\rm cool,all}$ represents the fraction of cooled gas that is either converted into stars or remains in the ISM phase at the redshift considered. Finally, $f_{*} = M_*/(M_* + M_{\rm cool})$ is the stellar mass fraction within the total mass of stars and ISM. Consistent with previous findings \citep[e.g.][]{kelly2022}, the total gas mass accreted onto each halo $M_{\rm acc}$ differs by no more than $15\%$ across all sample in this paper. Nevertheless, the accreted gas mass is systematically higher in the {\aur} runs than in {\apo}, suggesting that a smaller fraction of baryons is prevented from accreting in {\aur} and pointing to relatively weaker preventive feedback. This difference likely arises from variations in feedback implementation, with potential contributions from residual solver-dependent hydrodynamical effects.

\begin{figure}
    \includegraphics[width=0.5\textwidth]{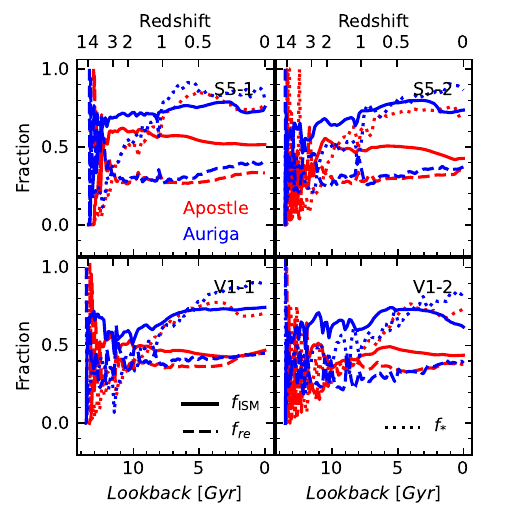}
    \centering
    \caption{Time evolution of the three factors entering the stellar mass decomposition for the {\aur} and {\apo} simulations. The solid lines show the cooling efficiency from the CGM to the ISM ($f_{\rm ISM}$) and the cold baryon retention fraction ($f_{\rm re}$), while the dot-dashed lines indicate the stellar mass fraction ($f_{*}$). The two simulations show significant differences in $f_{\rm ISM}$ but very similar values of $f_{\rm re}$.}
    \label{fig:decomHis}
\end{figure}

In Fig.~\ref{fig:decomHis}, we present the time evolution of the three factors in the two simulations. As shown by the solid lines in Fig.~\ref{fig:decomHis}, the two simulations exhibit significant differences in the $f_{\rm ISM}$ factor, indicating that the efficiency of gas cooling and condensation from the CGM into the ISM plays a dominant role in setting the final stellar mass. In particular, {\aur} provides roughly $40\%$ more gas available for star formation through this process compared to {\apo} at $z=0$.
Interestingly, $f_{\rm re}$ exhibits rather consistent behaviour in the two simulations, with a present-day value of approximately $0.4$. This result indicates that, despite the different simulation models, the galaxies retain cold baryons (stars plus star-forming gas) with nearly identical efficiencies once gas enters the star-forming phase. Further work will be required to investigate the origin of this possible universality.
The first two factors exhibit only weak time evolution after the early phases of galaxy evolution. In contrast, the star formation efficiency $f_{*}$ evolves with time and shows only moderate differences between the two simulations after $z=0.5$, suggesting that star formation within the ISM is self-regulated to a comparable level under the fiducial models, as indicated by the dot-dashed lines in Fig.~\ref{fig:decomHis}. Nevertheless, the {\aur} simulation still results in a somewhat larger amount of stellar mass formed at late times, consistent with our earlier finding that it hosts a younger stellar population. Our results indicate that variations in stellar mass between the simulations are largely driven by differences in the structure and properties of CGM, which regulate the supply of star-forming gas to the galaxy.

\subsection{Global Properties of Stellar Component} 
\label{subsec:props_star}

\begin{figure*}
    \includegraphics[width=0.9\textwidth]{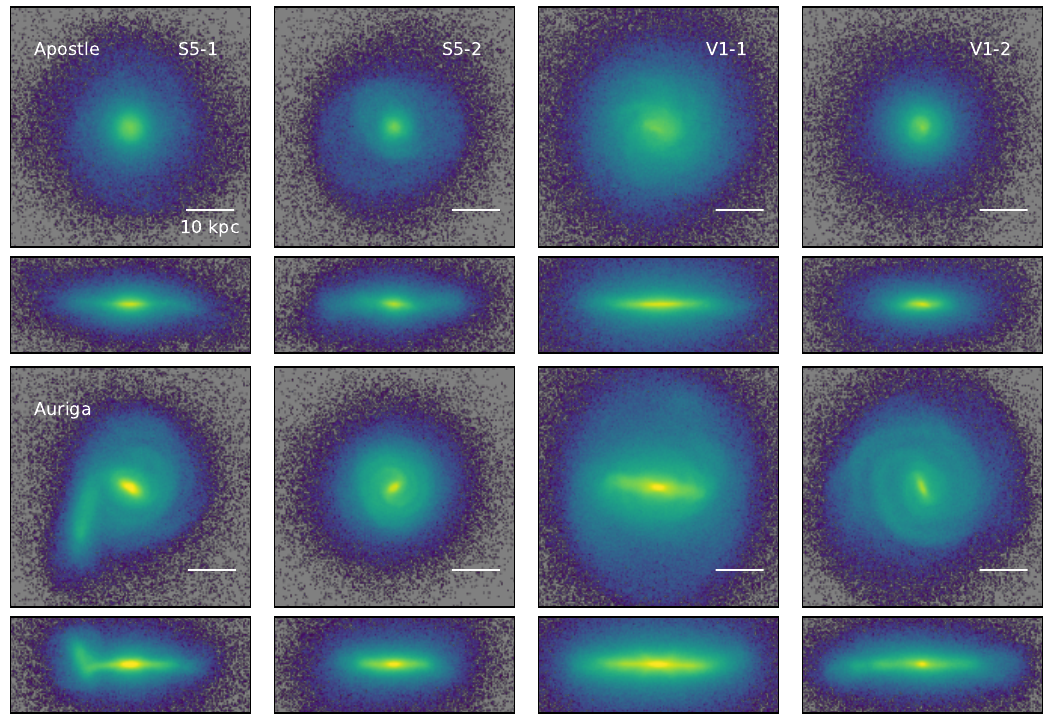}
    \centering
    \caption{The face-on and edge-on projected stellar surface density of galaxies at $z=0$. Top and bottom 2 rows are for counterparts in {\apo} and {\aur} simulations respectively. The projected box has dimensions of $50 \times 50 \times 20$ kpc along x, y, z directions.}
    \label{fig:projmap}
\end{figure*}

We now examine the stellar properties of these galaxies at $z=0$ following the comparison of their mass assembly histories. Given their disk-dominated nature (see \figref{fig:projmap}), the subsequent analysis is performed in a coordinate system aligned with the disk plane.

\begin{figure}
    \includegraphics[width=0.5\textwidth]{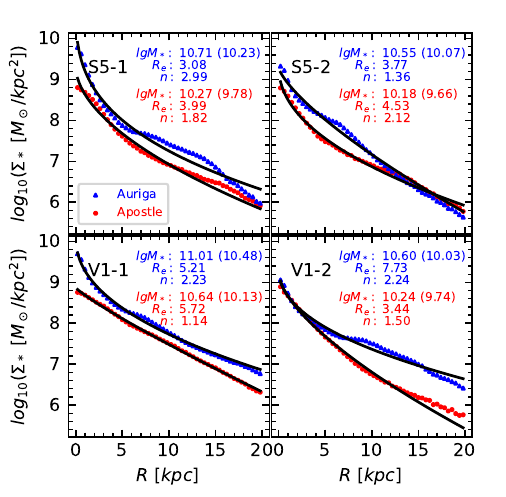}
    \centering
    \caption{Stellar surface density profiles at $z=0$. Each panel display results of galaxy counterparts (with name marked at top left) in the {\apo} (red curves) and {\aur} (blue curves) simulations. Black solid lines indicate the best-fit single {\sersic} profiles, whose parameters (total stellar mass $\lg M_*$, effective radius $R_e$, and {\sersic} index $n$) are listed in the top-right corner of each panel (color-coded to match the simulations). The stellar mass values in parentheses of $\lg M_*$ represent the actual summed masses of stellar particles within the analysis region $R \le$ 20 kpc and $|z| \le$ 5 kpc.}
    \label{fig:sdens}
\end{figure}

\figref{fig:sdens} presents the stellar surface density profiles, measured within $\pm 5$ kpc of the disk mid-plane along the z direction. We also fit each profile with a single {\sersic} model \citep{sersic1963, sersic1968}:
\begin{equation}  \label{eq:sersic}
    {\Sigma}(R) = \Sigma_e \exp{\{- b_{n} [(R/R_e)^{1/n}-1]\}}
\end{equation}
where $R_e$ is the effective radius enclosing half of the total masses, $\Sigma_e$ is the surface density at $R_e$ and $n$ is the {\sersic} index that quantifies the mass concentration. The normalization constant $b_n$ is a function of $n$ to ensure $R_e$ contains half of the total mass \citep{ciotti1991}. 

The fits are performed using weighted least squares, with each radial bin weighted by its statistical uncertainty. The uncertainties are estimated assuming Poisson noise in the particle counts within each bin ($\Delta\Sigma/\Sigma \propto 1/\sqrt{N}$). The best-fitting parameters — total stellar mass ($\log M_*$), effective radius ($R_e$), and Sérsic index ($n$) — are annotated in each panel, with colors denoting the simulation in \figref{fig:sdens}. {\aur} galaxies systematically exhibit higher surface densities than their {\apo} counterparts at all radii. This trend is consistent with their enhanced stellar mass growth histories in \figref{fig:evol_mass} and the projected surface densities in \figref{fig:projmap}. Moreover, three of the four {\aur} systems (excluding V1-2) have effective radii that are smaller by $10–20\%$ compared to their {\apo} counterparts. Similarly, three {\aur} galaxies (excluding S5-2) show larger Sérsic indices, indicating a more centrally concentrated stellar structure. These structural parameters are summarized in the last three columns of \tabref{tab:props}.

\begin{figure*}
    \includegraphics[width=0.48\textwidth]{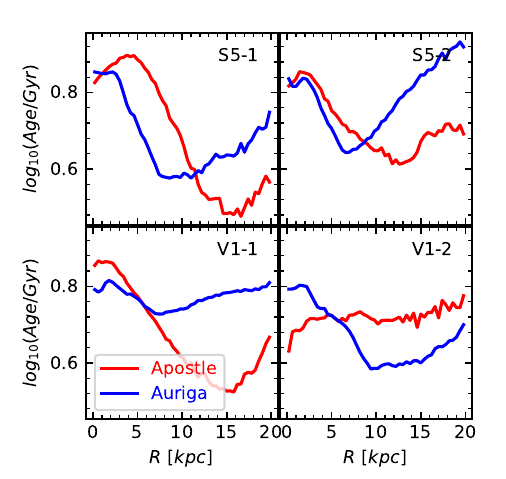}
    \includegraphics[width=0.48\textwidth]{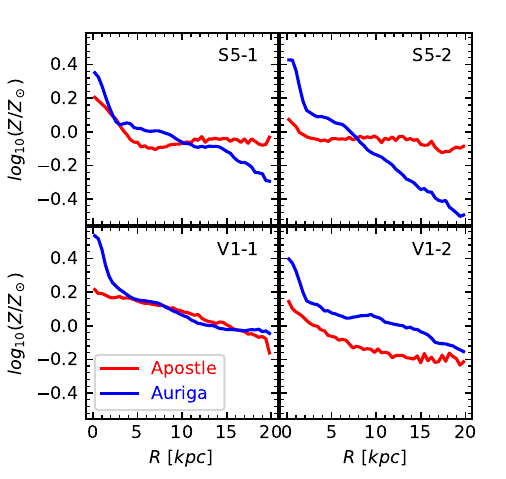}
    \centering
    \caption{Profiles of (mass-weighted) stellar age and metallicity for galaxies at $z=0$. Color coding is the same as that in \figref{fig:sdens}.}
    \label{fig:prof_age_Z_star}
\end{figure*}

We now examine the radial variations of the stellar populations through their mass-weighted age and metallicity profiles. \figref{fig:prof_age_Z_star} presents these profiles at $z=0$ in two simulations for the two simulations, measured over the same radial ranges as in \figref{fig:sdens}. The four {\aur} galaxies (blue curves) exhibit remarkably consistent trends. Their central regions ($R \lesssim 3$ kpc) are characterized by a nearly uniform mass-weighted stellar age of $\sim 6.3$ Gyr with super-solar mean metallicities ($\gtrsim 1.6 Z_\odot$). Even though the mean stellar age remains nearly constant in the inner regions, the metallicity drops steeply within the inner 3 kpc, and then declines smoothly to sub-solar values ($\sim 0.6 Z_\odot$) outward. The age profiles display characteristic U-shaped patterns, reaching minima at about 1--2 $R_{\rm hsm}$ before rising again in the outskirts. This upturn reflects the growing contribution of ex-situ stars in the outskirts, which are typically older. The trends of these two profiles in the {\apo} galaxies are similar to those in {\aur}, but the quantitative values differ significantly. As shown in the left panel of \figref{fig:sdens}, three galaxies (excluding V1-2) exhibit broadly similar U-shaped age profiles, but their turnovers occur at larger radii ($\sim 15$ kpc) than their {\aur} counterparts. 

As shown in the right panel of \figref{fig:sdens}, all {\apo} galaxies have systematically lower central metallicities than their {\aur} counterparts. In {\aur}, the high metallicity in the central regions can be attributed to the presence of a central gas clump that sustains ongoing star formation. A study based on TNG50 suggests that this central clump likely arises from numerical effects, compounded by the specific feedback implementation \citep{2025A&A...699A..12C}.

\begin{figure*}
    \includegraphics[width=0.9\textwidth]{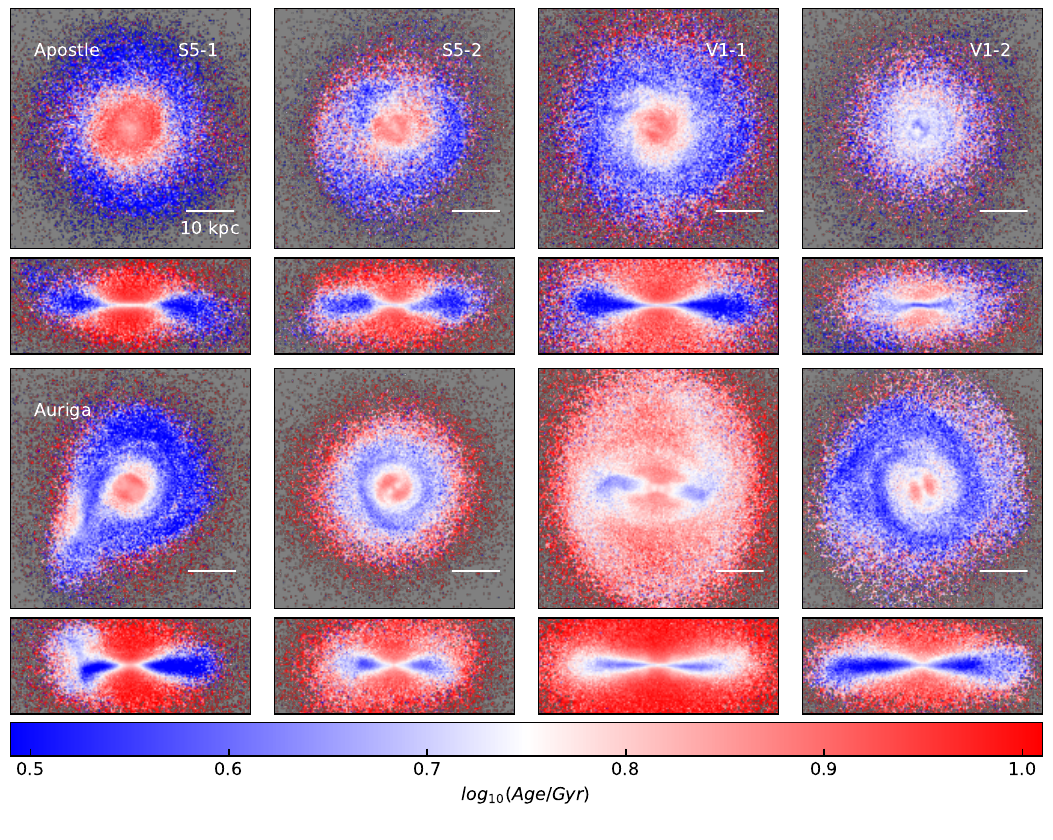}
    \centering
    \caption{The face-on and edge-on projected (mass-weighted) stellar age of galaxies at $z=0$. Plots are similar as \figref{fig:projmap}.}
    \label{fig:proj_age}
\end{figure*}

\begin{figure*}
    \includegraphics[width=0.9\textwidth]{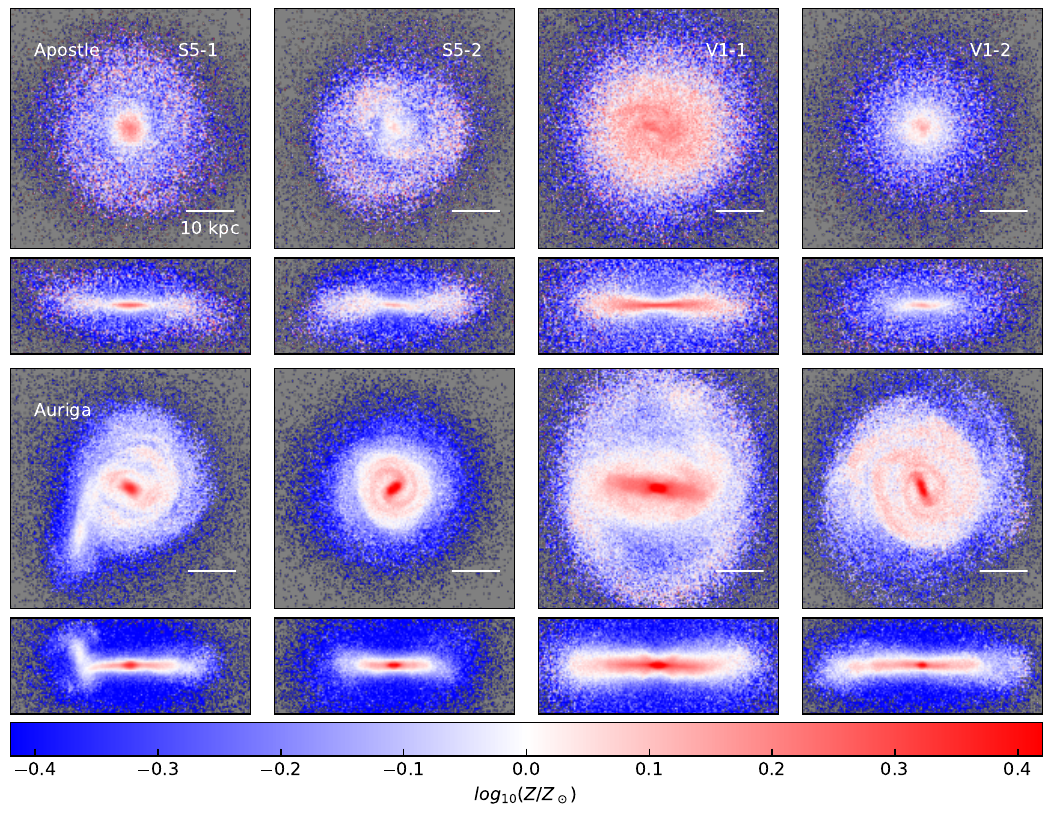}
    \centering
    \caption{The face-on and edge-on projected (mass-weighted) stellar metallicity of galaxies at $z=0$. Plots are similar as \figref{fig:proj_age}.}
    \label{fig:proj_Zstar}
\end{figure*}

\figref{fig:proj_age} and \figref{fig:proj_Zstar} present the two-dimensional projected distributions of mass-weighted stellar age and metallicity, computed over the same spatial regions as \figref{fig:projmap}.
The stellar populations exhibit a clear separation into three structurally distinct components:
(1) a compact central bulge, dominated by old, metal-rich stars with super-solar metallicity;
(2) an extended disk, where stars are systematically younger and have near-solar metallicity; and
(3) a diffuse stellar halo, composed primarily of old, metal-poor stars.
These distinct components naturally give rise to the radial features discussed above, including the U-shaped age profiles and the overall declining metallicity gradients. In the next section, we perform a quantitative structural decomposition to examine the physical properties of each component in detail.

\subsection{Morphology decomposition} \label{subsec:morph}

We begin by performing a morphological decomposition of each galaxy using its stellar surface density profile fitting. The bulge–disk decomposition method, widely used in observational studies of galaxies, describes galactic structures with simple parametric profiles. In this framework, the bulge component is modeled with a {\sersic} function (Eq. \ref{eq:sersic}), while the disk follows an exponential profile:
\begin{equation}
\Sigma(R) = \Sigma_0 \exp(-R/R_s),
\end{equation}
where $R_s$ denotes the disk scale length. This simple yet effective approach provides a good description of the main structural components of most disk-dominated galaxies, although additional components can be introduced for more complex systems.

\begin{figure*}
    \includegraphics[width=\textwidth]{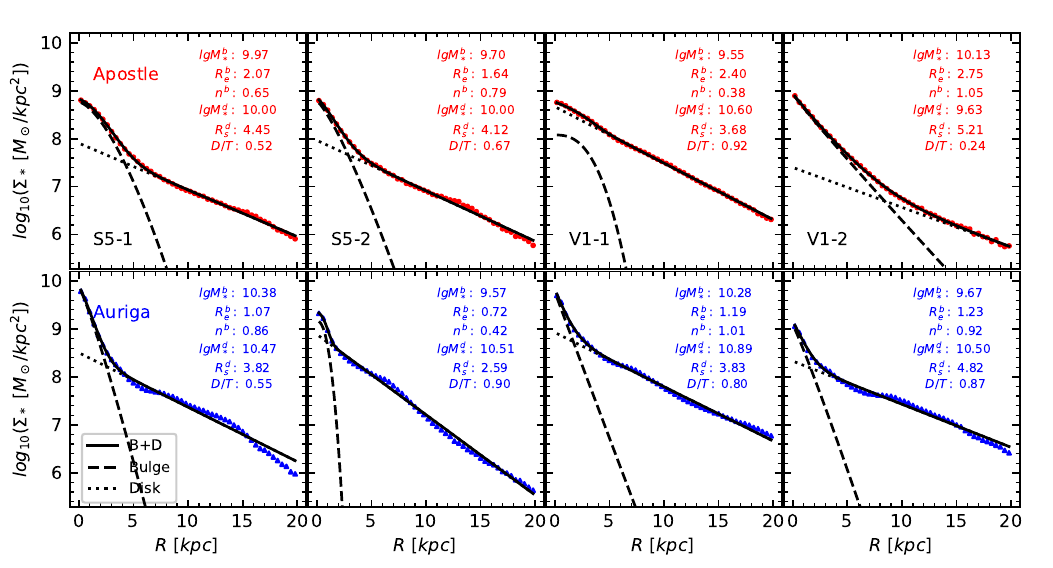}
    \centering
    \caption{Bulge/disk decomposition fit for galaxies at $z=0$. Plots are similar as \figref{fig:sdens}. Galaxies in {\apo} and {\aur} simulations are plotted seperately in top and bottom rows respectively. The dashed/dotted curves show bulge/disk components, with parameters denoted by superscripts \textit{b} and \textit{d} in the top-right texts.}
    \label{fig:bd_fit}
\end{figure*}

\figref{fig:bd_fit} shows the structural decomposition of the galaxies at $z=0$. Three of the four {\aur} systems (excluding V1-1) exhibit larger disk-to-total ratios ($D/T$), particularly S5-2 and V1-2 and they also feature somewhat shorter disk scale lengths ($R_s^d$, smaller by 10–40$\%$) compared with their {\apo} counterparts. In the V1-1 case, both simulations yield highly disk–dominated galaxies ($D/T \gtrsim 0.8$); however, the {\aur} galaxy develops a pronounced central light excess associated with its prominent bar. This morphological signature is clearly visible in the projected maps (\figref{fig:projmap}, \figref{fig:proj_age}, and \figref{fig:proj_Zstar}).

In addition to the profile–based decomposition, we perform a structural analysis using particle–level kinematics. Star particles are classified according to their orbital properties using the circularity parameter $\epsilon$ \citep{abadi2003}:
\begin{equation}
\epsilon \equiv j_z / j_{\rm circ}(E),
\end{equation}
where $j_z$ is the specific angular momentum component along the disk axis, and $j_{\rm circ}(E)$ is the angular momentum of a circular orbit with the same specific energy $E = \tfrac{1}{2}|\mathbf{v}|^2 + \psi$ as that star. Particles with $\epsilon>0$ are on prograde orbits, while those with $\epsilon<0$ are retrograde.

For efficiency, the gravitational potential is also calculated assuming spherical symmetry:
\begin{equation}
\psi(R) \approx -\frac{G M(<R)}{R} - \sum_{r_i>R} \frac{G m_i}{r_i},
\end{equation}
where the summation is over all particles outside radius $R$ in this halo.

\begin{figure*}
    \includegraphics[width=\textwidth]{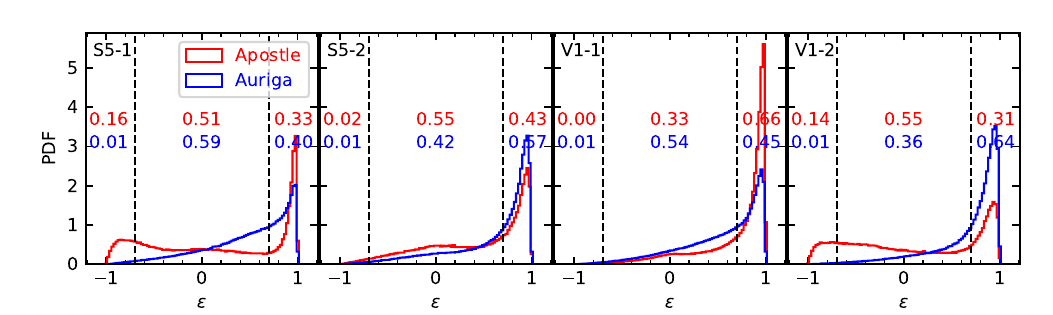}
    \centering
    \caption{Kinematic decomposition of galaxies at $z=0$. The mass-weighted probability distribution function of circularity $\epsilon$ for star particles. Red and blue colors are for simulations {\apo} and {\aur} respectively. Each column is for a matched galaxy pair from the two simulations, with name marked in top left corner of panels. The vertical dashed lines are for $\epsilon = \pm 0.7$, criteria for kinematic component decomposition. Value texts between these lines in the bottom row are mass fractions of stellar particles in corresponding $\epsilon$ bins: $\epsilon < -0.7$, $[-0.7, 0.7]$, $\epsilon > 0.7$, with color coding for simulations.}
    \label{fig:kinem_decomp}
\end{figure*}

\figref{fig:kinem_decomp} shows the stellar circularity distributions for each matched galaxy pair from the {\apo} (red) and {\aur} (blue) simulations at $z=0$. We adopt $\epsilon = 0.7$ as the threshold for kinematic decomposition, indicated by the vertical dashed lines. Only the coldest stellar components with $\epsilon > 0.7$ are classified as belonging to the kinematic disk.
All four {\aur} galaxies have kinematic disk-to-total ratios of $D/T \gtrsim 0.4$, while their {\apo} counterparts reach $D/T \gtrsim 0.3$. Although these values are systematically lower than those obtained from the profile-based decomposition—reflecting differences in the definition of components—the relative trends are preserved: three {\aur} systems (S5-1, S5-2, and V1-2) still exhibit larger kinematic $D/T$ values than their {\apo} analogs, with V1-1 showing the opposite behavior.
Notably, two {\apo} galaxies (S5-1 and V1-2) contain substantial retrograde components, with 16\% and 14\% of stars having $\epsilon < -0.7$, respectively. 

\begin{figure*}
    \includegraphics[width=0.48\textwidth]{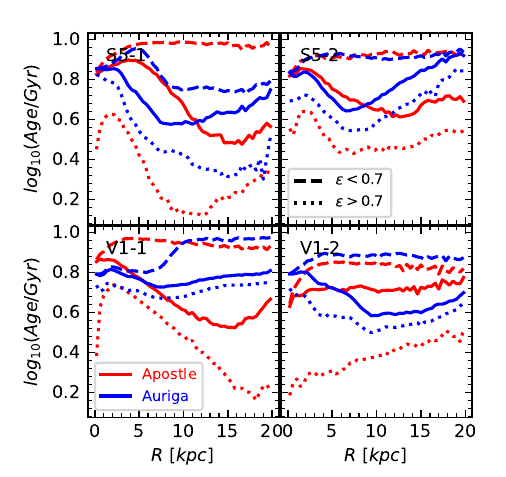}
    \includegraphics[width=0.48\textwidth]{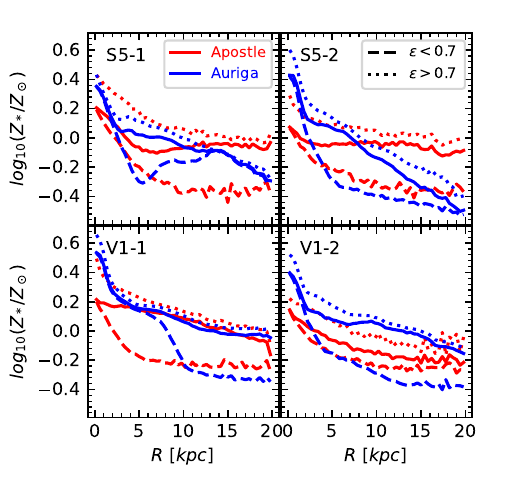}
    \centering
    \caption{Stellar age and metallicity profiles of individual kinematic components in galaxies at $z=0$. The solid lines represent the results for all stars. The dashed and dotted lines are for kinematic components with circularity $\epsilon < 0.7$ and $\epsilon > 0.7$ respectively. Other curves and color coding are the same as \figref{fig:prof_age_Z_star}.}
    \label{fig:prof_age_Z_kinem}
\end{figure*}

\figref{fig:prof_age_Z_kinem} shows the mass-weighted stellar age and metallicity profiles for the kinematically separated components, extending the analysis of \figref{fig:prof_age_Z_star}. The dashed and dotted lines represent the non-disk ($\epsilon < 0.7$) and disk ($\epsilon > 0.7$) populations, respectively.
The profiles reveal systematic differences between the two components in both simulations: disk populations consistently show younger ages and higher metallicities than the non-disk populations. The galaxies in {\aur} exhibit significantly older disk stars than their {\apo} counterparts, whereas the non-disk components have more comparable ages. This suggests that disks in {\aur} formed earlier than in {\apo}.
The right panel of \figref{fig:prof_age_Z_kinem} shows that, in the inner regions, the two components in {\aur} exhibit only small metallicity differences, whereas in {\apo} the disk component remains more metal-rich than the non-disk component at all radii. The metallicity contrast between the two components is therefore more pronounced in {\apo}, likely reflecting the later formation of the disk and less efficient metal mixing.

\begin{figure*}
    \includegraphics[width=\textwidth]{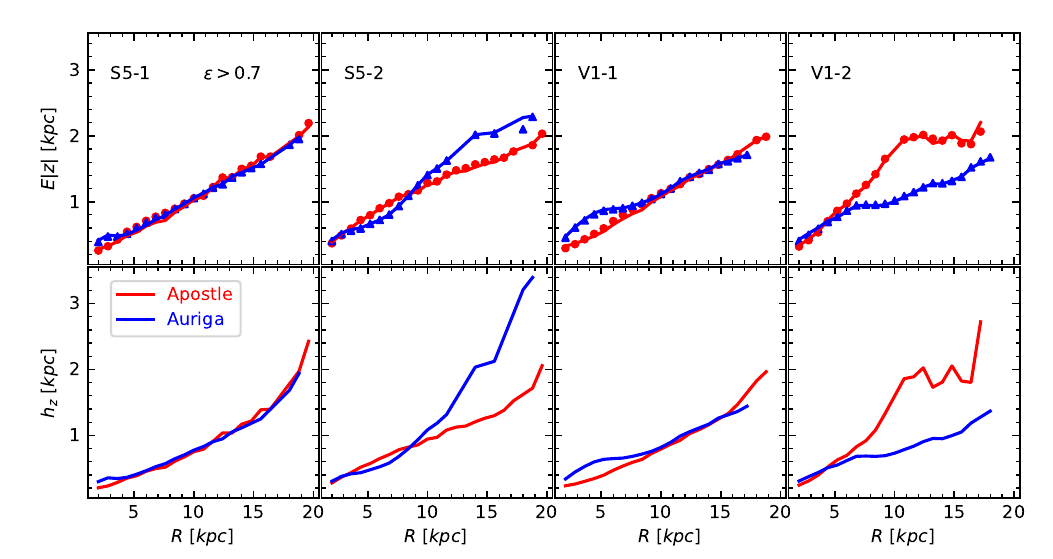}
    \centering
    \caption{Kinematic disk height as a function of radius. Only include kinematic disk stellar particles, with circularity $\epsilon > 0.7$. Top row: mass-weighted mean height $E|z|$ from \textit{true} vertical profile (blue triangles/red circles for measurements of {\aur}/{\apo} galaxies), and from best-fit {\sech} profile within the same region as the \textit{true} one (blue/red solid lines); bottom row: scale height $h_z$ of best-fit sech model as a function of radius.}
    \label{fig:hz_disk_kinem}
\end{figure*}

We further investigate the vertical structure of kinematic disk components by modeling their vertical density profiles with the {\sech} formulation \citep{vdkruit1981}:
\begin{equation} \label{eq:sech}
    \rho(z) = \rho_0~\text{sech}^2(\frac{z}{2 h_z})
\end{equation}
where $h_z$ is the scale height and $\rho_0$ is the midplane density at a given radius. This profile asymptotically approaches an exponential tail ($\rho \propto \exp{(-|z|/h_z})$) at large heights ($|z| \gg h_z$). The fitting is performed on vertical profiles extracted from annular radial bins, with the resulting scale height radial dependence shown in the bottom row of \figref{fig:hz_disk_kinem}. Radial bins without enough stellar particles for the fitting are excluded. To validate our modeling approach, we compare a model-independent measurement (mass-weighted mean height $E|z| \equiv (\Sigma m_i |z_i|)/(\Sigma m_i)$) of disk thickness (scattered points) in the top panels, calculated from both \textit{true} vertical profile of stellar particles and best-fit model one within the same z-range (solid lines, $E|z| = 2~\text{ln}2~h_z$ if integrating over the whole range). The excellent agreement between these measurements indicates that the {\sech} model provides a good description of the vertical distribution of disk stars across all radii, suggesting that the disks are close to vertical dynamical equilibrium. In approximate vertical equilibrium, the disk thickness can be described by the vertical Jeans equation. Within 10 kpc, the disks in the two simulations show similar thicknesses. This similarity may reflect broadly comparable inner gravitational potentials, consistent with the identical initial conditions adopted in the two simulations.


\subsection{Satellite galaxies} \label{subsec:satellites}

\begin{figure}
    \includegraphics[width=0.5\textwidth]{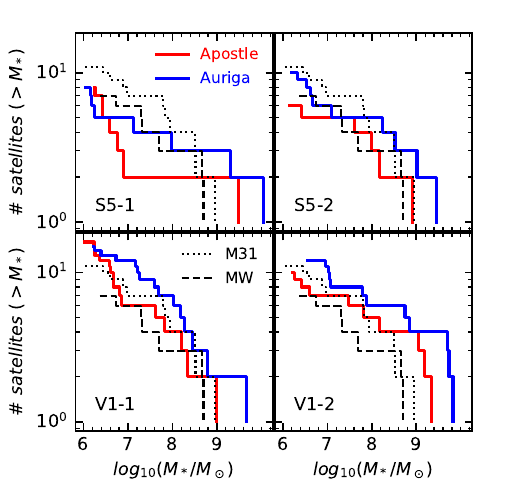}
    \centering
    \caption{Cumulative number of satellite galaxies as a function of stellar mass. Each panel corresponds to a halo pair (name shown in the bottom-left corner) from the {\apo} (red) and {\aur} (blue) simulations. The black dotted and dashed lines represent the observed satellite systems of M31 and the MW, respectively \citep{mcconnachie2012}.}
    \label{fig:cumn_satel_mstar}
\end{figure}

\begin{figure}
    \includegraphics[width=0.5\textwidth]{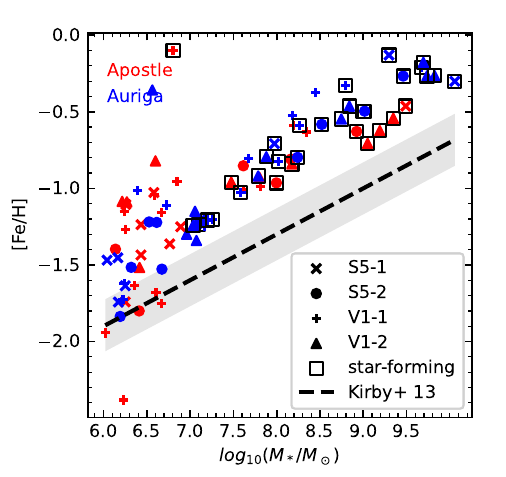}
    \centering
    \caption{Stellar mass–metallicity relation for satellite galaxies. Different symbols represent different halo pairs, with red and blue indicating {\apo} and {\aur}, respectively. Symbols enclosed by black squares highlight star-forming galaxies. The dashed line and grey shaded region show the mass–metallicity relation for MW and M31 dwarf satellites from \citet{kirby2013}, with the shaded region indicating the corresponding rms scatter of 0.17 dex.}
    \label{fig:mstar_mz_satel}
\end{figure}

In this section, we analyze the satellite galaxies surrounding our central Milky Way analogs in both the {\apo} and {\aur} simulations. We select all subhalos within a radius of 400 kpc from the central galaxies and exclude systems with stellar masses below $10^6M_\odot$.
Fig. \ref{fig:cumn_satel_mstar} presents the cumulative number of satellite galaxies for each central galaxy, compared with observational data from the Local Group (MW and M31) compiled by \citet{mcconnachie2012}.The satellite stellar mass functions differ systematically between the two simulations. However, in three of the four halo pairs, the total number of satellites with $M_* \ge 10^6 M_\odot$ is similar, broadly consistent with observations. All four {\aur} hosts contain more massive satellites than their {\apo} counterparts, with the most massive satellites being $\sim 0.3$ dex larger in stellar mass. A previous study \citep{2020MNRAS.492.5780R}, based on the original {\aur} and {\apo} simulations with different initial conditions, showed that the abundance of subhalos is reduced in hydrodynamical simulations relative to dark-matter-only runs, with a more pronounced reduction in {\aur}. This effect arises from the stronger tidal field of the more massive central galaxies in {\aur}, which enhances the disruption of subhalos. However, for dwarf-sized halos, galaxies formed in the {\aur} simulations tend to have significantly higher stellar masses \citep{yang2024}. The modest difference in the cumulative number of satellite galaxies may therefore result from the competition between these two opposing effects, although the most massive satellites in {\aur} still reach higher stellar masses.
\figref{fig:mstar_mz_satel} shows the stellar mass–metallicity relation of all satellite galaxies in both simulations, together with the observed relation for Milky Way and M31 dwarf galaxies (black dashed line; \citealt{kirby2013}). Both the {\apo} and {\aur} satellites exhibit systematically higher metallicities than observed systems at all stellar masses. In addition, the lowest-mass satellites with $M_* \sim 10^{6.5} M_{\odot}$ in the {\apo} simulations show unusually high average metallicities. This likely results from the absence of a metal diffusion model in {\apo}, which allows enriched gas particles to retain their metals and form stars with artificially high metallicities. Since these satellites are only marginally resolved—often consisting of just a few star particles—their mean metallicities are particularly sensitive to such effects.
In \figref{fig:mstar_mz_satel}, we classify a satellite as star-forming if its instantaneous star formation rate satisfies $\mathrm{SFR} > 0 M_{\odot}\mathrm{yr}^{-1}$ at $z=0$, and mark such systems with squares. In both simulations, most of the massive satellites with $M_* > 10^8 M_\odot$ remain star-forming, consistent with observations \citep[e.g.][]{2021ApJ...907...85M,2022ApJ...933...47C}. Both simulations also reproduce the observed trend that the quenched fraction of satellite galaxies increases toward lower stellar masses. However, they lack the population of faint blue satellites seen in observations \citep{2023MNRAS.519.4499P}, with this tension being more pronounced in the {\apo} simulations, as shown in \figref{fig:mstar_mz_satel}. These results suggest that more realistic treatments of baryonic physics may be required to better reproduce the observed satellite galaxy population.

\section{Conclusion and Discussion} \label{sec:discus}

In this study, we conducted a comprehensive comparison between the {\apo} and {\aur} simulations to investigate the impacts of different subgrid models and hydrodynamics solver on Milky Way analogs. By analyzing a sample of disk galaxies selected from both simulations with identical initial conditions but distinct hydrodynamical schemes and subgrid physics, our findings are summarized as follows:

\begin{enumerate}

    \item The dark matter mass assembly histories in both simulations are very similar, indicating that they are not sensitive to the differing hydrodynamical treatments and are therefore robust to baryonic processes in Milky Way-sized halo. In contrast, the growth of baryons shows a strong model dependence, with {\aur} yielding systematically higher baryon fractions in the halos across the cosmic time (Fig.~\ref{fig:evol_mass}).

    \item Our decomposition further indicates the stellar mass differences in Milky Way-sized halo between the simulations are largely driven by variations in the efficiency of CGM gas cooling into the ISM, highlighting the key role of CGM structure in regulating the supply of star-forming gas. At $z=0$, {\aur} provides roughly $40\%$ more gas available for star formation through this process compared to {\apo}. The retention fraction of baryons that have previously cooled, however, remains remarkably similar across the two simulations (Fig.~\ref{fig:decomHis}).

    \item Overall, both simulations produce well-defined disk structures. However, the Milky Way analogs in the {\aur} simulation exhibit higher stellar surface densities and total stellar masses than those in the {\apo} simulation. This more concentrated mass distribution leads to a shorter effective radius in the {\aur} galaxies (Fig.~\ref{fig:projmap}; Fig.~\ref{fig:sdens}). The mass-weighted radial profiles of stellar age and metallicity exhibit similar trends across the two simulations, albeit with quantitative differences. In particular, the stellar age profiles display a clear U-shaped radial profile, with older stars in the inner and outer regions and younger populations at intermediate radii (Fig.~\ref{fig:prof_age_Z_star}).
    
    \item The galaxies in the {\aur} simulation exhibit more prominent dynamical structures, such as bars and spiral arms (Fig.~\ref{fig:projmap}). These structures are also reflected in the mass-weighted projections of stellar age and metallicity. The stars located in these structures tend to have higher metallicities and younger ages. This result also indicates that a one dimensional projected radial profile is insufficient to fully describe the structural variations in the stellar age and metallicity distributions (Fig.~\ref{fig:proj_age}; Fig.~\ref{fig:proj_Zstar}).

    \item Both the surface density profile fitting and the kinematic decomposition indicate that galaxies in {\aur} generally have higher disk-to-total ratios ($D/T$), with three out of four cases showing larger values than their {\apo} counterparts. This trend is consistent with the angular momentum evolution reported in previous studies. However, strongly disk-dominated systems ($D/T>0.9$; V1-1 in {\apo} and S5-2 in {\aur}) are found in both simulations, although they do not correspond to the same host haloes. This suggests that the formation of disk structures depends on a complex interplay between halo assembly and baryonic processes  (Fig.~\ref{fig:bd_fit}; Fig.~\ref{fig:kinem_decomp}).

    \item In both simulations, the kinematically defined disk components are younger and more metal-rich, as expected. All {\aur} galaxies exhibit older disk stellar populations than their {\apo} counterparts, reflecting the relatively later formation of the kinematically cold disks in {\apo}. In the inner regions of the galaxies, the metallicity difference between the disk and non-disk components is small in {\aur}. In contrast, the disk component in {\apo} galaxies is about 0.2 dex more metal-rich than the non-disk component (Fig.~\ref{fig:prof_age_Z_kinem}).

    \item The vertical density profiles of disk stars are well fitted by a {\sech} model, suggesting that the disks are close to vertical dynamical equilibrium, with similar scale heights in the inner regions of the two simulations, both remaining below 1 kpc, while increasing gradually with radius due to disk flaring (Fig.~\ref{fig:hz_disk_kinem}).

    \item The satellite stellar mass functions differ significantly between the two simulations, with {\aur} producing systematically more massive satellites, and the most massive systems being $\sim 0.3$ dex higher in stellar mass. However, in three out of the four sample pairs, the number of satellite galaxies with stellar masses above $10^6 M_\odot$ is similar. In addition, the mass–metallicity relations predicted by the two simulations are very similar, although both are offset from the observational relation, with metallicities higher by about 0.25 dex at fixed stellar mass. Both simulations capture the observed increase in quenched fraction toward lower satellite stellar masses, yet they lack the faint blue satellite population found in observations (Fig.~\ref{fig:cumn_satel_mstar}; Fig.~\ref{fig:mstar_mz_satel}). 
    
\end{enumerate}

Our results indicate that galaxies produced by two simulations share similar properties in some aspects, but also show systematic differences in others, along with additional differences that are more complex and not strictly systematic. This suggests that the coupling between the baryonic physics processes and the assembly histories can lead to complex outcomes in Milky Way-analogs. Hence, relying on a few single scaling relations is insufficient to assess the validity of a model and a more systematic and comprehensive evaluation is required to capture the differences between models. While this comparative study has revealed systematic differences in galaxy properties between the {\aur} and {\apo} simulations, the exact causal relationships with their distinct subgrid physics implementations require further investigation.


\section*{Acknowledgements}

We acknowledge support from the National Natural Science Foundation of China (Grant No. 12588202) and the National Key Research and Development Program of China (Grant No. 2023YFB3002500). 

\bibliography{references}





\end{document}